\documentclass[%
reprint,
superscriptaddress,
 amsmath,amssymb,
 aps,
]{revtex4-1}

\usepackage{graphicx,setspace}
\usepackage{dcolumn}
\usepackage{bm}
\usepackage[mathlines]{lineno}
\usepackage[colorlinks=true,citecolor=blue,linkcolor=blue,anchorcolor=green, anchorcolor=green,urlcolor=blue]{hyperref}
\usepackage{mathrsfs}
\usepackage{xcolor}
\usepackage{float}
\usepackage{ulem}
\usepackage[utf8]{inputenc}

\begin{document}


\title{Stable circular orbits of spinning test particles around accelerating Kerr black hole}

\author{Ming Zhang}
\email{mingzhang@jxnu.edu.cn}
\affiliation{College of Physics and Communication Electronics, Jiangxi Normal University, Nanchang 330022, China}
\author{Jie Jiang}
\email{Corresponding author, jiejiang@mail.bnu.edu.cn}
\affiliation{Department of Physics, Beijing Normal University, Beijing 100875, China}

\date{\today}

\begin{abstract}
We investigate the stable circular orbits of the spinning test particles around the accelerating Kerr black hole on the equatorial plane. To this end, we first calculate the equations of motion and analyze the parameter space for the particles. We study the effect of the particle's spin and the black hole's acceleration on the conserved angular momentum, conserved energy and radius of the spinning test particle on the innermost stable circular orbit. We find that the effect of the particle's spin on the orbit parameters is almost linear, the effect of the black hole's acceleration on those parameters is nonlinear. We also explore the effects of the particle's spin and the black hole's acceleration on the periastron shift for the spinning particle in the nearly stable circular orbit. 

\end{abstract}


\maketitle


\section{Introduction}
For a massive particle revolving around  a central black hole on a circular orbit, there exists one orbit with minimal radius, which is named as the innermost stable circular orbit (ISCO). It is the last stable orbit on which the particle will not plunge into the black hole. The properties of ISCO convey  the information of the spacetime geometry and the central black body. The radius, angular momentum and energy of the particle on the ISCO in fact depend on the Killing vectors of the spacetime as well as hairs of the black hole. 

The binary black hole system with an  extreme mass ratio can be viewed as a test particle moving around a black hole. In view of this, the ISCO is the end stage of the relative circular motion of the system which at the same time emits gravitational waves \cite{Abbott:2016blz,Abbott:2016nmj}, and also the beginning of the inspiral motion. Another motivation to study the ISCO of the particle is to gain our recognition of the accretion disc  \cite{Page:1974he}.   

For a spherically symmetric black hole, without loss of generality, we usually study the particle on the equatorial plane. The ISCO of the massive particle for this kind of black hole is unique, irrespective of the direction of the particle's orbital angular momentum. Of course the characteristic quantities for this kind of ISCO are also definite. For instance, for a Schwarzschild black hole, the well-known result is that the radius of the ISCO for a massive particle is $6M$ with $M$ the mass of the black hole \cite{chandrasekhar1985mathematical}.

For the case where the central black hole is rotating, the ISCOs of the massive particles revolving around depend on the orbital angular momentum of the particles. That is, for the particles moving on the co-rotating orbit and on the counter-rotating orbit, their ISCOs are split, with distinct ISCO quantities. A typical example is the Kerr case. For massive particles revolving around  an extreme Kerr black hole with mass $M$, the radius of the ISCO for a prograde particle is $M$, whilst for a retrograde particle that radius, in contrast, becomes $9M$ \cite{bardeen1972jm}.

The cases mentioned above are for ideal test particles moving along the geodesics. To really achieve the aims of the ISCO investigation, we should also consider the properties of the test particles, as the particles we focus on in a realistic astrophysical process are extended objects owning internal structures. The internal structure of the test body reminds us that we should at least take the finite size effects at the dipole level into consideration, despite there are quadrupole and other higher multipole moments \cite{Steinhoff:2009tk}. In other words, we should at least consider the spin of the particle.

There are plenty of documentations on the studies of the ISCOs for not only the spin-less particles \cite{Liu:2017fjx,Chakraborty:2013kza,Pugliese:2011xn,Zahrani:2014rqa,Abdujabbarov:2009az,Delsate:2015ina,Isoyama:2014mja} but also the spinning ones \cite{Harms:2016ctx,Lukes-Gerakopoulos:2017vkj,Zhang:2017nhl,Hojman:1976kn}. As we have said above, the background spacetimes which were concerned are spherically symmetric and axially symmetric. (See the red more recent ones in \cite{Conde:2019juj,Toshmatov:2019bda}. And for the axially symmetric black hole, the ISCO investigation usually focuses on the equatorial plane. What we should further notice is that the southern hemispheres and the northern hemispheres of those axially symmetric black holes, e.g., the Kerr black hole, are also symmetric. In what follows, we will introduce our work about ISCO for the spinning test particle around the well-known accelerating Kerr black hole \cite{Griffiths:2005se} which is algebraically type-D \cite{Griffiths2009gravitation} and belongs to a larger class of Plebański-Demiański spacetime, including black hole parameters mass, electric charge, magnetic charge, NUT parameter, acceleration parameter, cosmological constant, and spin \cite{Griffiths:2005qp} (The effect of  the NUT parameter \cite{Bordo:2019rhu} in this spacetime may be investigated separately in future and we will not concentrate on it at present.). Due to the acceleration caused by the cosmic string \cite{Appels:2016uha}, the northern hemisphere and the southern hemisphere of the accelerating Kerr black hole are not identical. We will investigate the interplay of the particle's spin and the black hole's acceleration on the ISCO characteristic quantities (conserved angular momentum, conserved energy and radius) for the spinning particle on the equatorial plane of the accelerating Kerr black hole. As we will encounter an obstacle to analytically obtain the equations of motion for an orbit deviated from the equatorial plane, though it may be possible in the linear order of the particle's spin \cite{Mukherjee:2018zug} in a region far away from the acceleration horizon of the black hole (which makes the conformal factor normalized) \cite{Yao:2011ai}. The motion on the equatorial plane is adequate to reflect the effects of the black hole's acceleration on the system.

The generic elliptical orbit of a particle in the strong gravity region is not closed so that the particle will not return back to the initial point after an orbital period. This phenomenon is the well-known periastron shift (which is also named as periastron precession or periastron advance). The study of the periastron shift of a spinless particle around the Schwarzschild black hole can be seen in many textbooks, e.g., \cite{wald1984general}.  The periastron shifts for spinning binary black holes moving on quasi-circular orbits were computed using the effective-one-body formalism  in \cite{Hinderer:2013uwa} and using numerical-relativity simulations, the post-Newtonian approximation and theory of black hole perturbation in \cite{Tiec:2013twa}. Recently,  periastron shifts for a spinning  particle around the overcharging Reissner-Nordström spacetime and overspinning Kerr spacetime moving in nearly circular orbits  were studied in \cite{Mukherjee:2018zug}. In this paper, we will not only investigate the ISCO but also study the periastron shift for the spinning particle on the equatorial  nearly stable circular orbit of the accelerating Kerr black hole.

Based on this setup, the remaining parts of this paper are arranged as follows. In Sec. \ref{eos}, we will present the equations of motion as well as the constraints of the motion for a spinning test particle in the accelerating Kerr spacetime. In Sec. \ref{isco} we will show the interplay of the particle's spin and acceleration of the black hole on the characteristic quantities of the particle on the ISCO. In Sec. \ref{peri}, we will study the periastron shift for the spinning particle in the equatorial nearly stable circular orbit of the accelerating Kerr black hole. Sec. \ref{con} will be devoted to our closing remarks.

\section{Equations of motion for charged spinning test body in accelerating Kerr spacetime}\label{eos}
\subsection{Equations of motion}
The Mathisson-Papapetrou-Dixon (MPD) equations \cite{Papapetrou:1951pa,Dixon:1970zza}
\begin{equation}
\frac{D P^a}{D \tau }=-\frac{1}{2}R^a{}_{bcd}v^{b} S^{cd},
\end{equation}
\begin{equation}
\frac{D S^{\text{ab}}}{D \tau }=2P^{[a}v^{b]}
\end{equation}
should be used to model the motion of a spinning test particle. In the above equations, when the spin tensor $S^{ab}$ vanishes, the equations are reduced to describe the geodesic motion of a spinless particle. $P^{a}$ and $v^{a}$ are the four-momentum and four-velocity of the particle, respectively, and $\tau$ is the parameter along the trajectory of the particle.

To restrict the MPD equations to obtain the equations of motion for the spinning particle, we here use the Tulczyjew condition to ensure the conservation of the dynamical mass of the particle by choosing a consistent centre-of-mass, that is \cite{tulczyjew1959motion,dixon1964covariant,Dixon:1970zza}
\begin{equation}
S^{ab} P_b=0.
\end{equation}
The four-momentum of the particle defines the mass of the particle measured in the zero three-momentum frame by the relation \cite{Costa:2017kdr,Obukhov:2010kn}
\begin{equation}
 P^a  P_a=-\mathcal{M}^{2},
 \end{equation}
which at the same time means that the normalized dynamical four-momentum of the particle should be
\begin{equation}
u^a\equiv \frac{P^a}{\mathcal{M}}.
\end{equation}
The contract of the four-velocity $v^a$ and the four-momentum $P^a$ gives the other mass of the particle measured in the zero three-velocity frame \cite{Costa:2017kdr,Obukhov:2010kn}, as  $P_av^a=-m$.

As the magnitude $S$ of the spin for the particle is conserved, we have \cite{Wald:1972sz}
\begin{equation}
S^{ab}S_{ab}=2 S^2.
\end{equation}

Utilizing the above conditions, one can get the difference between the four-velocity and the normalized four-momentum for the spinning particle as \cite{Costa:2017kdr,Obukhov:2010kn,Hojman:1976kn,Lukes-Gerakopoulos:2017cru}
\begin{equation}\label{relationuv}
v^a=N\left(u^a+\frac{2 S^{ab} u^c R_{bcde} S^{de}}{S^{bc} R_{bcde} S^{de}+4 \mathcal{M}^2}\right),
\end{equation}
where 
\begin{equation}
N\equiv \frac{m}{\mathcal{M}}\nonumber.
\end{equation}

For the spinless particle, the mass $\mathcal{M}$ and $m$ are identical. According to the Tulczyjew SSC, we know that  $\mathcal{M}$ is invariable along the worldline of the test particle. However, the mass $m$ is variable. We have $\mathcal{M}=m+\mathcal{O}(S^{2})$, which means that, at the linear order of spin of the particle, the mass $\mathcal{M}$ and $m$ cannot be  differentiated \cite{Ruangsri:2015cvg}.     Under the reparametrization of the orbital parameter $\tau$, we thus can fix the parameter to satisfy $v^{a}u_{a}=-1$ \cite{Tanaka:1996ht,Saijo:1998mn,Mukherjee:2018zug} so that $N=1$.

The spacetime background we study in this article is the accelerating Kerr black hole, which  can be described by the line element \cite{Anabalon:2018qfv}
\begin{equation}
\begin{aligned}
ds^2=&\frac {dt^2 \left(\Delta -a^2 P \sin ^2 \theta \right)}{\alpha ^2 \Sigma  \Omega ^2}+\frac{dr^2 \Sigma }{\Delta  \Omega ^2}+\frac{d\theta^2 \Sigma }{P \Omega ^2}\\&+\frac{d\phi^2 \sin^2\theta\left[P \left(a^2+r^2\right)^2-a^2 \Delta  \sin ^2\theta\right]}{K^2 \Sigma  \Omega ^2}\\&-\frac{2 a dt d\phi \sin ^2\theta \left[P \left(a^2+r^2\right)-\Delta \right]}{\alpha  K \Sigma  \Omega ^2},
\end{aligned}
\end{equation}
where 
\begin{eqnarray}
\Sigma~ &=&r^2+a^2 \cos ^2\theta,\nonumber~\\ \Delta &=&(1-A^{2}r^{2})(r^{2}-2 Mr+a^2),\nonumber\\ \Omega&=&1+A r \cos \theta,\nonumber\\P&=&1+2 A M \cos \theta +a^2 A^2 \cos ^2\theta\nonumber ,\\ \alpha&=&\frac{\sqrt{\left(1-a^2 A^2\right) \left(a^2 A^2+1\right)}}{a^2 A^2+1}\nonumber.
\end{eqnarray}
In this metric, $M, a$ are the mass and angular momentum per mass of the black hole, just like its Kerr counterpart. $A$ is the acceleration parameter of the black hole, $K$ produces the conical deficits on the north or the south pole. $\alpha$ is a rescaling parameter which makes the Killing vector related to the time coordinate be normalized at conformal infinity. The conformal boundary of the spacetime is determined by the conformal factor $\Omega$. The conical deficits at the two poles are 
\begin{equation}
\delta_{\pm}=2\pi\left(1-\frac{P_{\pm}}{K}\right),
\end{equation}
with
\begin{equation}
P_{\pm}=1\pm 2MA+a^{2}A^{2}\nonumber
\end{equation}
corresponding to $P(\theta=0)$ and  $P(\theta=\pi)$, respectively. The tensions on the two poles are
\begin{equation}
\mu_{\pm}=\frac{\delta_{\pm}}{8\pi}.
\end{equation}

Choosing the normalized tetrad 
\begin{equation}
e_a^{(0)}dx^a =\frac{1}{\Omega}\sqrt{\frac{\Delta }{\Sigma }} \left(\frac{dt}{\alpha}-a \sin^{2}\theta \frac{d\phi}{K} \right),\label{tetrad}
\end{equation}
\begin{equation}
e_a^{(1)}dx^a =\frac{1}{\Omega}\sqrt{\frac{\Sigma }{\Delta }}dr,\label{tetrad2}
\end{equation}\begin{equation}
e_a^{(2)}dx^a = \frac{1}{\Omega}\sqrt{\frac{\Sigma}{P} }d\theta, \label{tetrad3}
\end{equation}\begin{equation}
e_a^{(3)}dx^a =\frac{\sin \theta}{\Omega}\sqrt{\frac{P}{\Sigma}} \left[-\frac{a dt}{\alpha}+\left(a^2+r^2\right)\frac{d\phi}{K}\right],\label{tetrad4}
\end{equation}
the accelerating Kerr metric can be expressed as
\begin{equation}
g_{ab}=\eta_{(i)(j)}e_{a}^{(i)}e_{b}^{(j)},
\end{equation}
with $\eta_{(i)(j)}=\text{diag}(-1,1,1,1)$.

It has been verified that the conserved quantity of the spinning particle in a spacetime relating with the Killing vector $\xi^a$ is \cite{Dixon:1970zza}
\begin{equation}
C_{\xi }=-\frac{1}{2} S^{ab} \nabla _b\xi _a+\xi _a P^a.
\end{equation}
For the accelerating Kerr spacetime, it is not difficult to see that $\xi_{t}\equiv \left(\frac{\partial }{\partial t}\right)^a$ and $\xi_{\phi}\equiv \left(\frac{\partial }{\partial \phi }\right)^a$ are two Killing vectors that produce two conserved quantities—the conserved energy $e$ and the conserved angular momentum $j$—as
\begin{equation}
-C_{\xi_{t}}=e=\frac{1}{2 \mathcal{M}}S^{tb} \nabla _b\xi _t-\xi _t u^t,
\end{equation}
\begin{equation}\label{conj}
C_{\xi_{\phi}}=j=-\frac{1}{2 \mathcal{M}}S^{\phi b} \nabla _b\xi_\phi+u^\phi \xi_\phi.
\end{equation}

The spin tensor of the spinning particle is related to the particle's spin vector via the relation 
 \begin{equation}
S^{(c) (d)}=\mathcal{M}\varepsilon^{(c)(d)}_{}{}_{(a) (b)}u^{(a)}s^{(b)},
\end{equation}
where $\varepsilon_{(a)(b)(c)(d)}$ is the completely antisymmetric tensor and $\varepsilon_{(0)(1)(2)(3)}=1$.

As we focus on the motions of the particle on the equatorial plane, we have $v^{(2)}=u^{(2)}=0$. We can set 
\begin{equation}
s^{(2)}=-s,\, s^{(0)}=s^{(1)}=s^{(3)}=0,
\end{equation}
where $s>0$ and $s<0$ mean that the spin directions are parallel and antiparallel to the spin of the black hole, respectively. Then we can obtain the non-vanishing components of the spin tensor as
\begin{eqnarray}
S^{(0)(1)}&=&-\mathcal{M} s u^{(3)},\\
S^{(0)(3)}&=&\mathcal{M} s u^{(1)},\\
S^{(1)(3)}&=&\mathcal{M} s u^{(0)}.
\end{eqnarray}

Thus the conserved energy and conserved angular momentum can be further expressed as
\begin{eqnarray}
e=\frac{\sqrt{\Delta}}{\alpha r}u^{(0)}+\frac{a r+s \left(A^2 M r^2-A^2 r^3+M\right)}{\alpha  r^2}u^{(3)},
\end{eqnarray}
\begin{equation}\begin{aligned}
j=&\left(-\frac{s \left(A^2 r^2-1\right) \left(a^2 \sqrt{\Delta }-2 M r+r^2\right)}{\Delta  K r}+\frac{a \sqrt{\Delta }}{K r}\right)u^{(0)}\\&+\frac{a^2 r+a s \left(A^2 M r^2-A^2 r^3+M+r\right)+r^3}{K r^2}u^{(3)},
\end{aligned}\end{equation}
with which one can further get the components of the normalized four-momentum for the spinning particle as
\begin{equation}\begin{aligned}
u^{(0)}=&\frac{\alpha  e r \left(a^2 r+a s \left(A^2 M r^2-A^2 r^3+M+r\right)+r^3\right)}{\sqrt{\Delta }\mathcal{X}}\\&-\frac{j K r \left(a r+s \left(A^2 M r^2-A^2 r^3+M\right)\right)}{\sqrt{\Delta }\mathcal{X}},
\end{aligned}\end{equation}
\begin{eqnarray}
u^{(2)}&=&0,\\
u^{(3)}&=&\frac{r^2 \left[j K-\alpha  e (a+s)\right]}{\mathcal{X}},\\
u^{(1)}&=&\sigma \sqrt{-1+(u^{(0)})^2  - (u^{(3)})^2}=\sigma  \sqrt{V},\label{ref}
\end{eqnarray}
where $\mathcal{X}=r^3 \left(A^2 s^2+1\right)-M s^2 \left(A^2 r^2+1\right)$, the particle is outgoing for $\sigma=1$ and ingoing for $\sigma=-1$, and $V$ is the radial effective potential for the particle.

Using (\ref{relationuv}), we have the four-velocity of the spinning particle  as
\begin{eqnarray}
v^{(0)}&=&\left(1+\frac{3 Ms^{2}\left(u^{(3)}\right)^{2}}{\mathcal{P}_{2}}\right)u^{(0)},\\
v^{(1)}&=&\left(1+\frac{3 Ms^{2}\left(u^{(3)}\right)^{2}}{\mathcal{P}_{2}}\right)u^{(1)},\\
v^{(3)}&=&\left(1+\frac{3 Ms^{2}\left(1+\left(u^{(3)}\right)^{2}\right)}{\mathcal{P}_{2}}\right)u^{(3)},
\end{eqnarray}
where
\begin{equation}\begin{aligned}
\mathcal{P}_2=r^3-M^{2}s^{2}-3M^{2}s^{2}\left(u^{(3)}\right)^{2}.\nonumber
\end{aligned}\end{equation}

On the other hand, the four-velocity for the charged spinning particle is
\begin{equation}
v^a=\left(\frac{\text{dt}}{\text{d$\tau $}},\,\frac{\text{dr}}{\text{d$\tau $}},\, 0,\,\frac{\text{d$\phi $}}{\text{d$\tau $}}\right),
\end{equation}
Realizing that 
\begin{equation}
v^{(\alpha)}=e^{(\alpha)}_{a}v^{a},
\end{equation}
we have
\begin{eqnarray}
v^{(0)}&=&\frac{1}{\Omega}\sqrt{\frac{\Delta }{\Sigma }} \left(\frac{1}{\alpha}\frac{dt}{d\tau}-\frac{a\sin ^2\theta}{K}\frac{d\phi }{d\tau }\right)\label{velocityone},\\
v^{(1)}&=&\frac{1}{\Omega} \sqrt{\frac{\Sigma }{\Delta }}\frac{dr}{d\tau},\\
v^{(3)}&=&\frac{\sin \theta}{\Omega}\sqrt{\frac{P}{\Sigma}}\left[\frac{a^2+r^2}{K}\frac{d\phi}{d\tau }-\frac{a}{\alpha}\cdot\frac{dt}{d\tau }\right],\label{velocitytwo}
\end{eqnarray}
which give
\begin{equation}
\frac{dt}{d\tau}=\frac{2 \alpha  \Omega  \left[ \left(a^2+r^2\right) \sqrt{P\Sigma}v^{(0)}+a  \sqrt{P\Delta }  v^{(3)}\right]}{2 r^2\sqrt{\Delta P } },
\end{equation}
\begin{equation}\label{rtau}
\frac{dr}{d\tau}=\Omega\sqrt{\frac{\Delta}{\Sigma}}v^{(1)},
\end{equation}
\begin{equation}\label{phitau}
\frac{d\phi}{d\tau}=\frac{K \sqrt{\Sigma } \Omega  \left(a \sqrt{\Delta } P v^{(0)}+\Delta  \sqrt{P} v^{(3)}\right)}{\Delta  P r^2}.
\end{equation}
These are exactly the equations of motion for the spinning particle revolving around the accelerating Kerr black hole on the equatorial plane.

\subsection{Value space of key parameters}
The spin of the  spinning test particle is positive irrespective of the reference frames we choose, we then have \cite{Wald:1972sz}
\begin{equation}
  s=\frac{S}{m}\lesssim r_0\ll r_+ =M+\sqrt{M^2-a^2}\leqslant 2M,
\end{equation}
where $r_0$ is the size of the particle and $r_+$ is the event horizon radius of the accelerating Kerr black hole which can be obtained by solving $\Delta=0$.

The time-like condition of the four-velocity gives
\begin{equation}
\mathcal{Y}v^\mu v_\mu\sim  -r^{24} \left(1-A^2 r^2\right) \left(a^2+r (r-2 M)\right)<0,
\end{equation}
where
\begin{equation}
\mathcal{Y}=\mathcal{F}(r,A,s,e,j,K,M)^2>0.\nonumber
\end{equation}
For simplicity, we will not show the tedious expression of $\mathcal{F}$.  Then we should keep
\begin{equation}
A^2 <\frac{1}{4M^2}.
\end{equation}

The conformal factor $\Omega$ gives the conformal boundary
\begin{equation}
r_\Omega =\left|-\frac{1}{A\cos\theta}\right|,
\end{equation}
which is also named as the accelerated horizon. For $\theta=\pi/2$, the accelerated horizon locates at the spatial infinity \cite{Anabalon:2018qfv}. We in this paper only consider that $r_\Omega\gg r_+$ for the case $\theta\neq\pi/2$, which yields that $A\ll 1$.

The forward-in-time condition which forbids the movement of the particle back in time \cite{Grib:2013hxa,wald1984general} demands
\begin{equation}
\begin{aligned}
\frac{dt}{d\tau}=&\frac{M s (a e-j K)+a r (e (a+s)-j K)+e r^3}{r^3}\\&\times \frac{a \left(\sqrt{a^2+r (r-2 M)}+a\right)+r^2}{a^2+r (r-2 M)}+\mathcal{O}({A^{2}})\\ >&0,
\end{aligned}
\end{equation}
which gives
\begin{equation}
j<\frac{e \left(a^2 r+a s (M+r)+r^3\right)}{K (a r+M s)}+\mathcal{O}({A^{2}}).
\end{equation}
We can see that, if $K\to\infty$, only a particle with negative conserved angular momentum (i.e., the retrograde particle) complies with the forward-in-time condition.

 \begin{figure*}[!tbp] 
   \centering
   \includegraphics[width=7in]{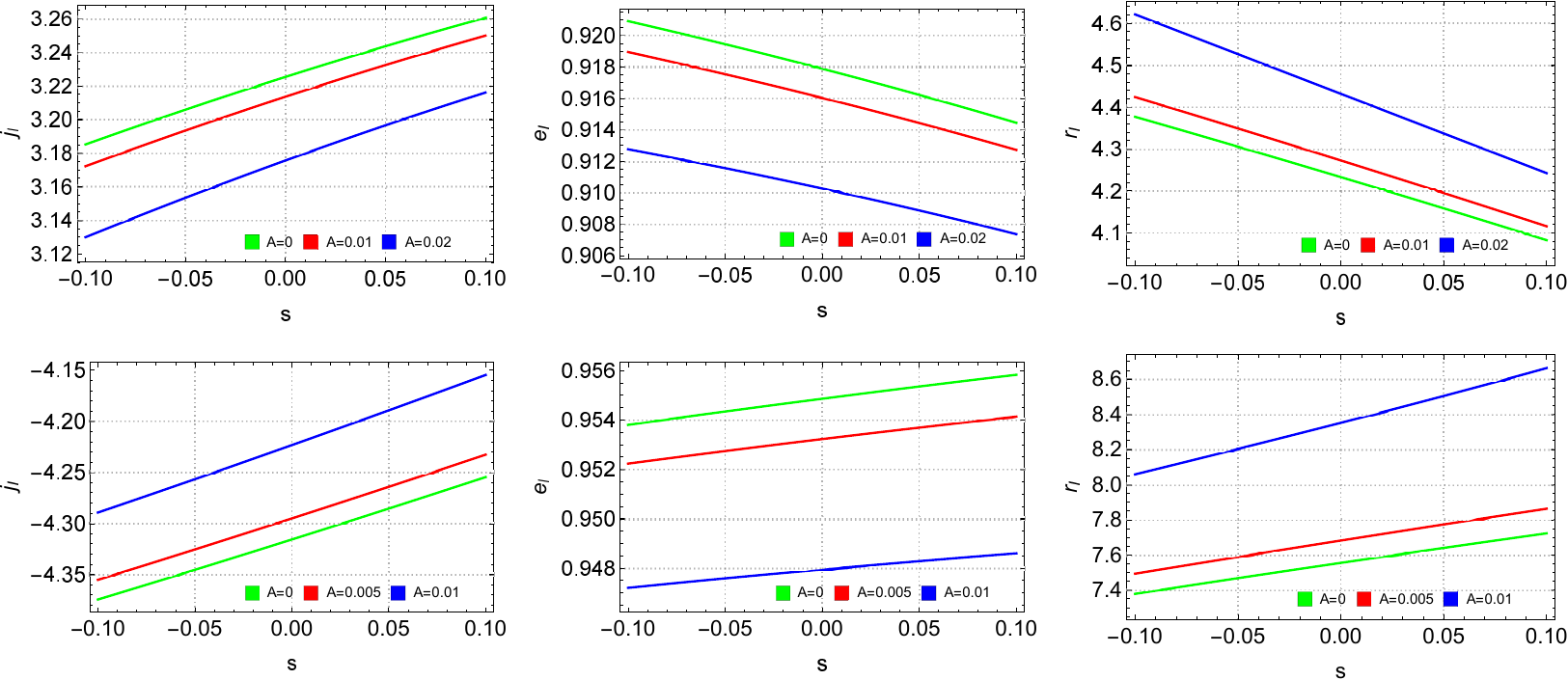}
   \caption{The ISCO parameters—conserved angular momentum $j_{I}$, conserved energy $e_{I}$ and radius $r_{I}$—in terms of the spin of the particle with the upper panel for the prograde orbit and the lower panel for the retrograde orbit. Other parameters are set to be $a=0.5,\,K=0.9,\,M=1$.}
   \label{p1}
\end{figure*}

 \begin{figure*}[!tp] 
   \centering
   \includegraphics[width=7in]{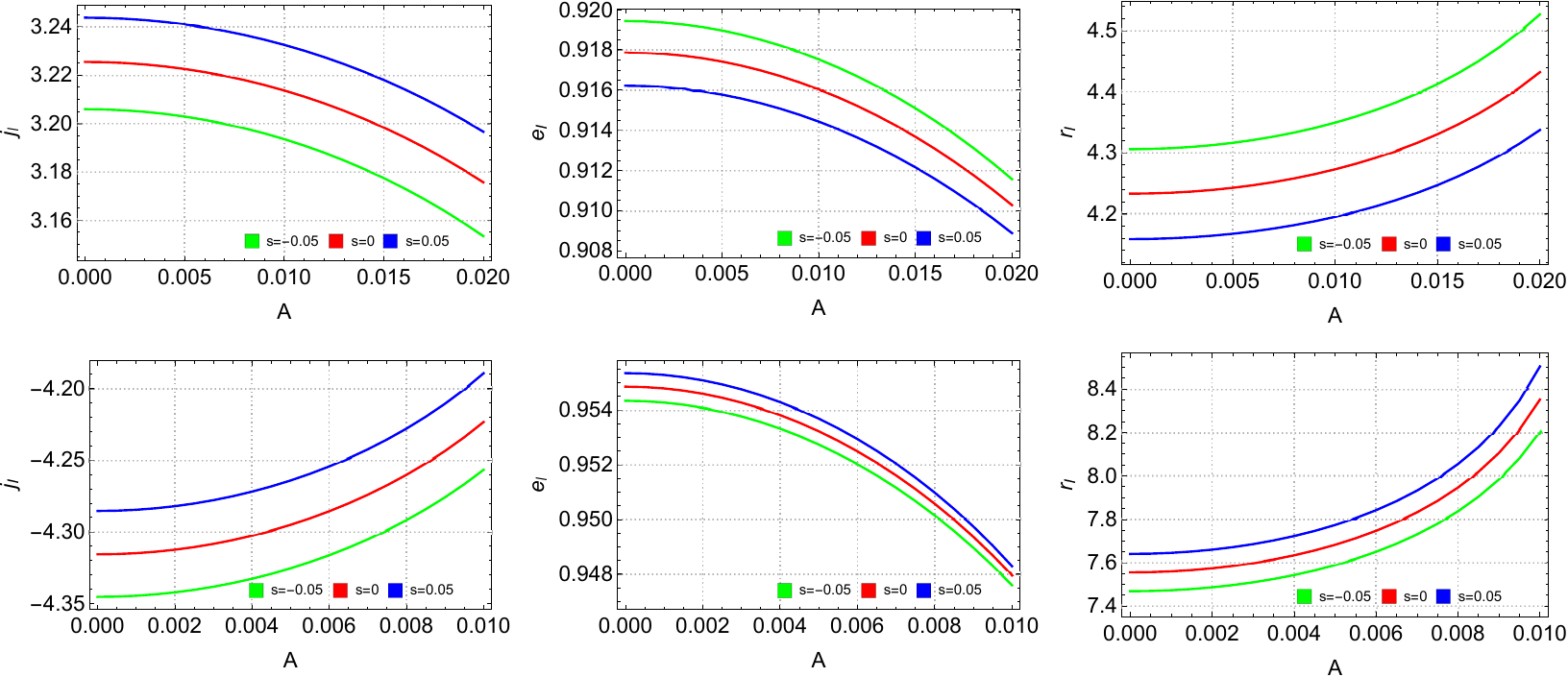}
   \caption{The ISCO parameters—conserved angular momentum $j_{I}$, conserved energy $e_{I}$ and radius $r_{I}$—in terms of the acceleration of the black hole with the upper panel for the prograde orbit and the lower panel for the retrograde orbit. Other parameters are set to be $a=0.5,\,K=0.9,\,M=1$.}
   \label{p2}
\end{figure*}

\section{The ISCO of the spinning particle around the accelerating Kerr black hole}\label{isco}
To obtain the ISCO parameters of the spinning particle, i.e., the conserved angular momentum $j_I$, the conserved energy $e_I$ and the radius $r_I$, we should set
\begin{equation}\label{conone}
\frac{dr}{d\tau}=0,
\end{equation}
\begin{equation}\label{contwo}
\frac{d^2 r}{d{\tau}^2}=0,
\end{equation}
\begin{equation}\label{conthree}
\frac{d^2 V}{dr^2}=0,
\end{equation}
where $V$ is the radial effective potential of the spinning particle defined in (\ref{ref}). The first condition restricts that the radial velocity of the particle vanishes; the second condition makes the radial acceleration absent; the third condition further ensures that the location of the particle is at the inflection point of the effective potential.

The effective potential of the particle is related to the radial velocity as
\begin{equation}
V\sim\left(\frac{dr}{d\tau}\right)^2\sim \left(v^{(1)}\right)^2\sim \left(u^{(1)}\right)^2.
\end{equation}
Explicitly, We have
\begin{equation}\label{squramo}
(u^{(1)})^2=\frac{\kappa e^2 +\beta e + \gamma}{\mathcal{X}^2 \Delta},
\end{equation}
where
\begin{equation}\begin{aligned}
\kappa = \frac{r^2 Z_1 \left(1-a^2 A^2\right)}{a^2 A^2+1},
\end{aligned}\end{equation}
\begin{equation}\begin{aligned}
\beta =-\frac{2 j K r^2 Z_4 \sqrt{1-a^4 A^4}}{a^2 A^2+1},
\end{aligned}\end{equation}
\begin{equation}\begin{aligned}
\gamma = a^2 Z_7+2 a j^2 K^2 r^3 s \left(A^2 M r^2-A^2 r^3+M\right)+r Z_9,
\end{aligned}\end{equation}
with 
\begin{equation}\begin{aligned}
Z_{1}=&a^4 A^2 r^4-2 a M r^3 s \left(A^2 r^2-3\right)+r^6\\&+2 a^3 M r s \left(A^2 r^2+1\right)+r^3 s^2 \left(A^2 r^2-1\right) (r-2 M)\\&+a^2 Z_2,\nonumber
\end{aligned}\end{equation}
\begin{equation}\begin{aligned}
Z_{2}=r^3 \left(M \left(2-2 A^2 r^2\right)+A^2 r^3+r\right)+s^2 Z_3,\nonumber
\end{aligned}\end{equation}
\begin{equation}\begin{aligned}
Z_{3}=M \left(2 r-2 A^4 r^5\right)+A^4 r^6+\left(A^2 M r^2+M\right)^2-A^2 r^4,\nonumber
\end{aligned}\end{equation}
\begin{equation}\begin{aligned}
Z_{4}=&a^3 A^2 r^4 - r^3 (r + M (-3 + A^2 r^2)) s\\&+a^2 r s \left(2 M \left(A^2 r^2+1\right)-A^2 r^3\right)+a Z_5,\nonumber
\end{aligned}\end{equation}
\begin{equation}\begin{aligned}
Z_{5}=&M^2 \left(A^2 r^2 s+s\right)^2+A^2 r^4 \left(s^2 \left(A^2 r^2-1\right)+r^2\right)\\&+M Z_6,\nonumber
\end{aligned}\end{equation}
\begin{equation}\begin{aligned}
Z_{6}=-2 A^2 r^5+s^2 \left(-2 A^4 r^5-A^2 r^3+r\right)+2 r^3,\nonumber
\end{aligned}\end{equation}
\begin{equation}\begin{aligned}
Z_{7}=&s^4 \left(A^2 r^2-1\right) \left(A^2 M r^2+M\right)^2\\&-2 M r^3 s^2 \left(A^4 r^4-1\right) \left(A^2 s^2+1\right)\\&+r^6 Z_8,\nonumber
\end{aligned}\end{equation}
\begin{equation}\begin{aligned}
Z_{8}=&A^6 r^2 s^4+A^2 \left(j^2 K^2+r^2-2 s^2\right)-1\\&-A^4 \left(s^4-2 r^2 s^2\right),\nonumber
\end{aligned}\end{equation}
\begin{equation}\begin{aligned}
Z_{9}=&-2 M^3 s^4 \left(A^2 r^2-1\right) \left(A^2 r^2+1\right)^2\\&+r^7 \left(A^2 r^2-1\right) \left(A^2 s^2+1\right)^2\\&+j^2 K^2 r^5 \left(A^4 r^2 s^2+A^2 r^2-1\right)\\&+M^2 r s^2 Z_{10} \left(A^2 r^2+1\right)-2 M r^4 Z_{11},\nonumber
\end{aligned}\end{equation}
\begin{equation}\begin{aligned}
Z_{10}=&j^2 K^2 \left(A^2 r^2+1\right)\\&+\left(A^2 r^2-1\right) \left(5 A^2 r^2 s^2+4 r^2+s^2\right),\nonumber
\end{aligned}\end{equation}
\begin{equation}\begin{aligned}
Z_{11}=&\left(A^2 r^2-1\right) \left(A^2 s^2+1\right) \left(s^2 \left(2 A^2 r^2+1\right)+r^2\right)\\&+j^2 K^2 \left(A^4 r^2 s^2+A^2 \left(r^2+s^2\right)-1\right).\nonumber
\end{aligned}\end{equation}
The radial effective potential can be defined as the minimum allowable energy at position $r$ for the spinning particle \cite{Conde:2019juj}, so we have
\begin{equation}\label{veff}
V\sim\frac{-\beta+\sqrt{\beta ^2-4 \kappa  \gamma }}{2 \kappa }.
\end{equation}
Then we can obtain the ISCO parameters $r_I,~j_I ,~e_I$ by using Eqs. (\ref{conone}), (\ref{contwo}), (\ref{conthree}) and (\ref{veff}). We can have an analytical solution for $A=0\,,s=0$, whist we have to resort to numerical calculations for other cases. We have shown the results in Figs. \ref{p1} and \ref{p2}, where the interplay of the particle's spin and the acceleration of the black hole are displayed.

 \begin{figure*}[!tp] 
   \centering
   \includegraphics[width=3.3in]{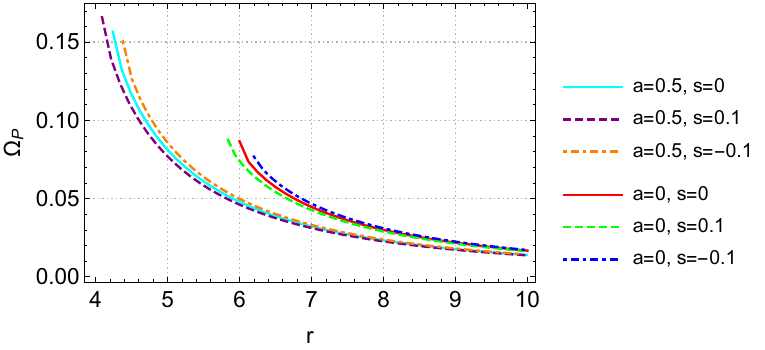}
   \includegraphics[width=3.3in]{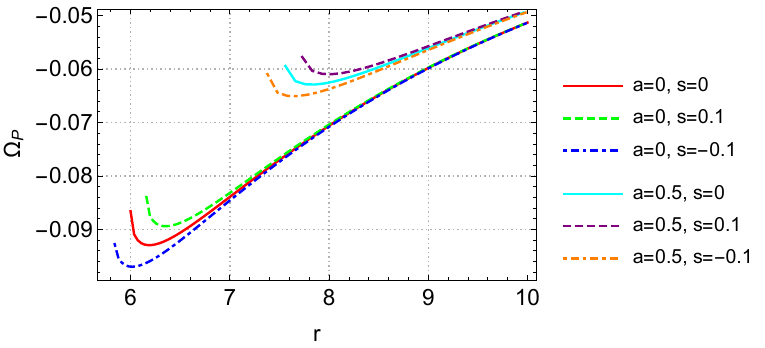}
   \caption{Variations of the periastron shifts with respect to the radial coordinate with $M=1, K=0.9, $ and $A=0$. The left one is for the prograde orbit and the right one is for the retrograde orbit.}
   \label{p3}
\end{figure*}

 \begin{figure*}[!tp] 
   \centering
   \includegraphics[width=3.3in]{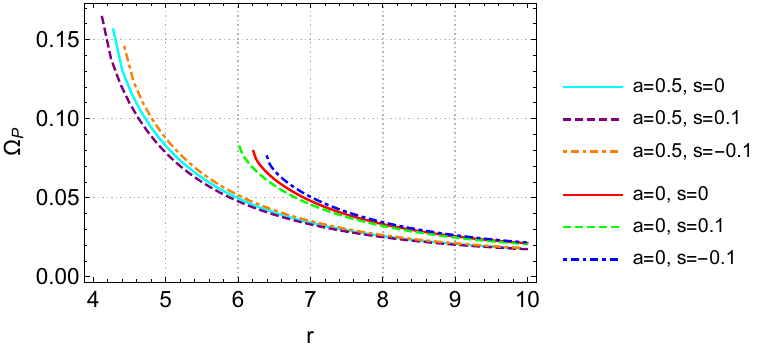}
   \includegraphics[width=3.3in]{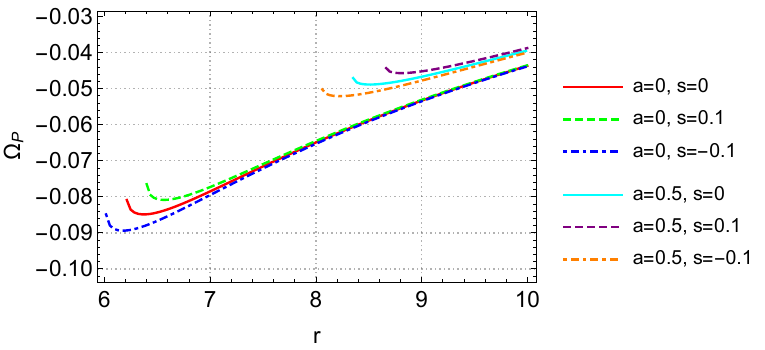}
   \caption{Variations of the periastron shifts with respect to the radial coordinate with $M=1, K=0.9, $ and $A=0.01$. The left one is for the prograde orbit and the right one is for the retrograde orbit.}
   \label{p4}
\end{figure*}

 \begin{figure*}[!tp] 
   \centering
   \includegraphics[width=3.3in]{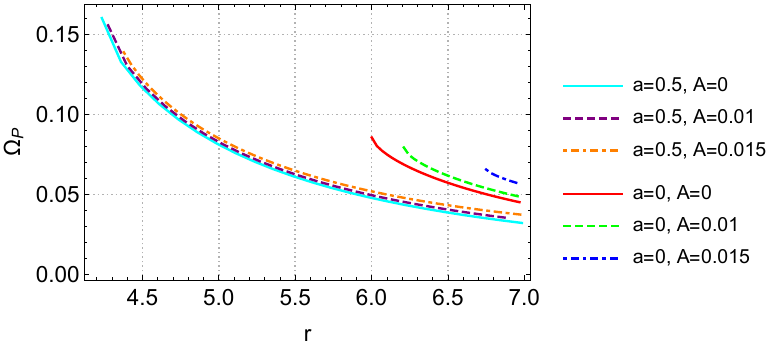}
   \includegraphics[width=3.3in]{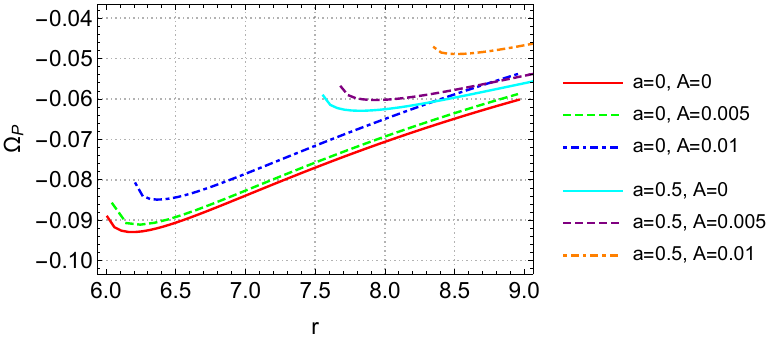}
   \caption{Variations of the periastron shifts with respect to the radial coordinate with $M=1, K=0.9, $ and $s=0$. The left one is for the prograde orbit and the right one is for the retrograde orbit.}
   \label{p5}
\end{figure*}

 \begin{figure*}[!tp] 
   \centering
   \includegraphics[width=3.3in]{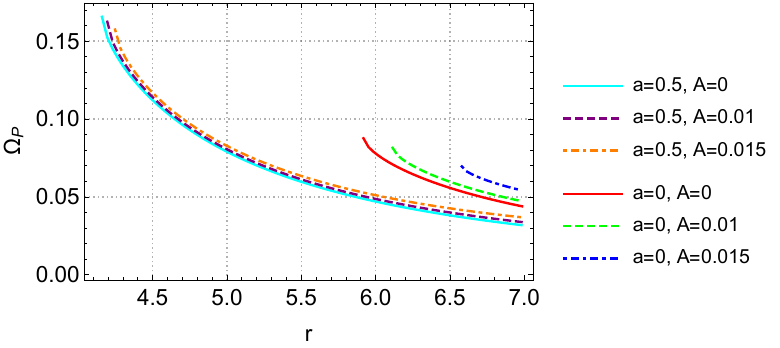}
   \includegraphics[width=3.3in]{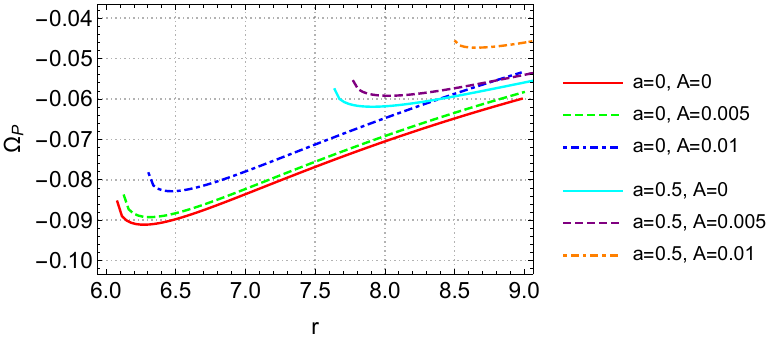}
   \caption{Variations of the periastron shifts with respect to the radial coordinate with $M=1, K=0.9, $ and $s=0.05$. The left one is for the prograde orbit and the right one is for the retrograde orbit.}
   \label{p6}
\end{figure*}

In Fig. \ref{p1}, we have shown the variations of the particle's conserved angular momentum, conserved energy and radius on ISCO with respect to the particle's spin. As $s\ll1$, we choose $-0.1<s<0.1$. The time-like condition and the forward-in-time condition of the particle are checked. We choose three different values of the acceleration $A$. The result is obvious, only the conserved angular momentum increases with the particle's spin on the prograde ISCO, whilst all the three parameters—$r_I,~j_I ,~e_I$—increase with the spin for the retrograde ISCO.

In Fig. \ref{p2}, we have shown the variations of the ISCO parameters with respect to the acceleration $A$ of the black hole.  As we need to have an acceleration horizon far away from the ISCO and the event horizon of the accelerating Kerr black hole, we have chosen $0<A<0.02$ for the prograde orbit and $0<A<0.01$ for the retrograde orbit. We see that both the radii of the prograde and the retrograde ISCOs increase with the acceleration; both the particles' conserved energy on the two kinds of ISCOs decrease with the acceleration.  Differently, we have decreasing $j_{I}$ for the prograde orbit and increasing $j_{I}$ for the retrograde orbit when the acceleration of the black hole increases.

The parameter $K$ is related to the average deficit of the black hole \cite{Gregory:2019dtq,Zhang:2019vpf}. 
By a transformation
\begin{equation}\label{rphi}
\phi\to\phi^{\prime}=K\phi,
\end{equation}
and then combining it with Eq. (\ref{conj}), we can see that $K$ relates with the conserved angular momentum of the particle on ISCO via a simple rescaling relation, and it will not affect the conserved energy and the radius of the particle on the ISCO. So different values of $K$ does not qualitatively change our results in Figs. \ref{p1}, \ref{p2}.

\section{Periastron shift of spinning particle around accelerating Kerr black hole}\label{peri}
If the spinning particle on the stable circular orbit with radius $r_{0}$ is slightly displaced on the radial direction, it will move in a harmonic form with a frequency 
\begin{equation}
\Omega_{r}=\frac{1}{2}\left(\frac{d^{2}V}{dr^{2}}\right)_{r=r_{0}}.
\end{equation}
Else, the angular frequency for the stable circular orbit is
\begin{equation}
\Omega_{\phi}=\frac{d\phi}{d\tau}.
\end{equation}
In the Minkowski spacetime limit, we have $\Omega_{r}=\Omega_{\phi}$; however, in the strong gravity region near the black hole, the radial oscillatory  frequency is not equal to the angular frequency. The difference between them is defined as the periastron shift 
\begin{equation}
\Omega_{P}=\Omega_{\phi}-\Omega_{r}.
\end{equation}
We numerically calculate the periastron shifts for the spinning test particle in the equatorial nearly stable circular orbits around the accelerating Kerr black hole and show our obtained results in Figs. \ref{p3}-\ref{p6}. 

Preliminarily, Comparing Fig. \ref{p3} with Fig \ref{p4}, we can see that variations of periastron shifts  with respect to the radial coordinate for the spinning test particle share similar characteristics in the background of both the accelerating Kerr black hole (including its Schwarzschild limit) and the Kerr black hole (including its Schwarzschild limit). (Note that $K=1$ in the Kerr/Schwarzschild cases, but we set $K=0.9$ here. As shown in Eq. \ref{rphi}, this does not change the results qualitatively.) Also, comparing Fig. \ref{p5} with Fig \ref{p6}, we can know that variations of periastron shifts for the test particle with respect to the radial coordinate share similar characteristics, whether the spin of the particle is considered or not.

By further analyzing, we get some other features as follows:

(1) The leftmost point on every curve corresponds to the ISCO and all the periastron shifts go asymptotically to zero when the radial coordinate extends to spatial infinity. The periastron shift decreases monotonically with respect to the radial coordinate for spinning or spinless particle on prograde orbit under the background of a black hole endowed with acceleration/angular momentum or without acceleration/ angular momentum. In contrast, the negative periastron shift for the particle on retrograde orbit first decreases to the minimum at a position near the ISCO and then increases monotonically  with respect to the radial coordinate. 

(2) According to Fig. \ref{p3} and Fig. \ref{p4}, we know that the periastron shift of the particle on the prograde orbit will be boosted by the particle spin antiparallel to the black hole spin and will be weakened by the particle spin parallel to the black hole spin; the periastron shift of the particle on the retrograde orbit will be boosted  by the particle spin parallel to the black hole spin and will be weakened by the particle spin antiparallel to the black hole spin. 

(3) According to Fig. \ref{p5} and Fig. \ref{p6}, we know that the acceleration of the black hole has an effect of increasing the periastron shifts for the particles both on the prograde orbit and the retrograde orbit.

(4) The periastron shifts for the particle on the prograde ISCOs increase with the spin of the particle and decrease with the acceleration of the black hole; the periastron shifts for the particle on the retrograde ISCOs increase with the spin of the particle and increase with the acceleration of the black hole.

(5) For the particle on the retrograde orbit, the minimum value of the periastron shift always increases with the acceleration and the angular momentum of the black hole as well as the spin of the particle.

\section{Closing remarks}\label{con}
We investigated the innermost stable circular orbits of the spinning test particles in the accelerating Kerr spacetime. Using MPD equations with proper supplement conditions, we gave the equations of motion for the spinning particle. After analyzing the parameter space, we further studied the interplay of the particle's spin, the black hole's acceleration on the angular momentum, energy and the radius of the particle on the ISCO, which was elucidated in detail in Sec. \ref{isco}. On the whole, we found that the effect of the particle's spin on the ISCO parameters is almost linear, whereas the effect of the black hole's acceleration is non-linear, even we have, starting from physically reasonable conditions, set them to be much less than unity. This can be explained by analyzing the radial effective potential of the spinning particle which can be expanded as
\begin{equation}\begin{aligned}\label{4815}
V\approx&2 a^2 e^2 M r^5-a^2 r^6-4 a e j K M r^5+2 j^2 K^2 M r^5\\&+a^2 e^2 r^6+e^2 r^8-j^2 K^2 r^6+2 M r^7-r^8\\&+s\left(2 a^3 e^2 M r^3-4 a^2 e j K M r^3+2 a j^2 K^2 M r^3\right.\\&\left.+6 a e^2 M r^5-6 e j K M r^5+2 e j K r^6\right)\\&+A^{2}\left(-4 a^4 e^2 M r^5-a^4 e^2 r^6+4 a^3 e j K M r^5\right.\\ &\left. -2 a^3 e j K r^6-2 a^2 e^2 M r^7+a^2 j^2 K^2 r^6\right.\\&\left.+a^2 r^8++4 a e j K M r^7-2 j^2 K^2 M r^7\right.\\&\left.-a^2 e^2 r^8-2 a e j K r^8+j^2 K^2 r^8-2 M r^9+r^{10}\right)\\&+\mathcal{O}(s)+\mathcal{O}(A^{2})+\mathcal{O}(sA^{2}).
\end{aligned}\end{equation}
We see that there is a linear term of the spin $s$ but there is only a quadratic term of the acceleration $A$. Moreover, we notice that in some literature (e.g., Ref. \cite{Chen:2016tmr}), the metric of the accelerating Kerr black hole is different from ours as they take $A$ by $-A$. Considering the characteristics of the effective potential (\ref{4815}) for the spinning particle, we know that this sign difference does not qualitatively change our conclusion. We here do not observe the degeneracy of the orbits for the particle, which was found in \cite{Zhang:2018eau}. This can also be roughly explained by the characteristics of the effective potential, as there is not term about $sA$. The rescaling factor $\alpha$ does not exist in the accelerating Kerr metric in early literature \cite{Griffiths:2005se}, and it was added in recent works \cite{Appels:2016uha,Appels:2017xoe,Abbasvandi:2018vsh,Abbasvandi:2019vfz} for constructing a consistent thermodynamics. Anyway, by expanding this dubious factor, we have
\begin{equation}
\alpha\approx 1-a^{2}A^{2}+\mathcal{O}(A^{2}).
\end{equation}
We see that it does only shift the conserved energy of the particle on the ISCO at a scale of $A^{2}$, which can be neglected. So $\alpha$ does not affect our conclusion quantitatively.

We also investigated the periastron shift for the spinning particle in the equatorial nearly stable circular orbit of the accelerating Kerr black hole. We mainly found that, the periastron shift of spinning particle on the prograde orbit decreases with the particle's spin and the periastron shift of the spinning particle on the retrograde orbit increases with particle's spin; the periastron shift of the spinning particle increases with the acceleration of the black hole. These results are also suitable for the Kerr black hole case and Schwarzschild black hole case.

\section*{Acknowledgements}
Jie Jiang  is supported by the National Natural Science Foundation of China (Grant No.11675015). Ming Zhang is supported by the Initial Research Foundation of Jiangxi Normal University with Grant No. 12020023.

\end{document}